# Magnetic field driven novel phase transitions in EuTiO$_3$

P. Pappas[1], M. Calamiotou[2], M. Polentarutti[3], G Bais[3], A. Bussmann-Holder[4], and E. Liarokapis[1]

[1]Department of Physics, National Technical University of Athens, Athens 15780, Greece
[2]Section of Condensed Matter Physics, Physics Department, National and Kapodistrian University of Athens, GR-15784 Athens, Greece
[3]Elettra-Sincrotrone Trieste S.C.p.A. Strada Statale 14, km 163.5, Basovizza, 34149 Trieste, Italy
[4]Max-Planck-Institute for Solid State Research, Heisenbergstr. 1, D-70569 Stuttgart, Germany

**Abstract**

The influence of an external static magnetic field (up to 480mT) on the structural properties of EuTiO$_3$ (ETO) polycrystalline samples was examined by powder XRD at the Elettra synchrotron facilities in the temperature range 100-300K. While the cubic to tetragonal structural phase transition temperature in this magnetic field range remains almost unaffected, significant lattice effects appear at two characteristic temperatures (~200 K and ~250 K), which become more pronounced at a critical threshold field. At ~200 K a change in the sign of magnetostriction is detected attributed to a modification of the local magnetic properties from intrinsic ferromagnetism to intrinsic antiferromagnetism. These data are a clear indication that strong spin-lattice interactions govern also the high temperature phase of ETO and trigger the appearance of magnetic domain formation and novel phase transitions.

Magneto-dielectrics and multiferroics are in the focus of present research activities due to their multifunctional properties and the resulting large potential of applications. However, a substantial drawback is either their rare occurrence or the insufficient coupling strength between polar and magnetic properties. A promising candidate to realize the desired material properties is represented by the perovskite EuTiO$_3$ (ETO) where magnetism stems from the 4f$^7$ electrons of Eu$^{2+}$ at the A site in the perovskite ABO$_3$ lattice, which leaves the transition metal B site in the d$^0$ configuration thus enabling the occurrence of ferroelectricity. Polar properties are a consequence of the dynamical covalence between the unoccupied (d$^0$) transition metal d-states and the oxygen ion p states.[1] In spite of the early discovery of ETO, it was not of any interest for more than half a century since "only" antiferromagnetism was observed below 5K[2,3] and the cubic symmetry believed to be given at all temperatures. A certain breakthrough was achieved when *T. Katsufuji and H. Takagi* discovered that the onset of magnetism was accompanied by a dielectric anomaly, which in turn was associated with a soft long wavelength transverse optic mode that extrapolates to zero far below T=0K, signaling a tendency towards ferroelectricity.[4] Besides of the fact that this soft optic mode is analogous to the one observed in SrTiO$_3$ (STO)[5,6] ETO and STO also share the cubic perovskite structure with almost the same lattice constant a=3.905 Å, valence of Eu, Sr 2+, Ti 4+, and space group Pm-3m. From a detailed analysis of the dynamical properties of ETO within the polarizability model the cubic to tetragonal phase transition (Pm-3m→I/4mcm) was predicted to occur near 300K, and subsequently verified experimentally.[1,7] This phase transition is – in analogy to STO – driven by a soft zone boundary acoustic mode. The order parameter of the transition is the oxygen ion octahedral rotation angle φ.[7] In spite of the apparent resemblance between ETO and STO significant differences between them stem from their local double well potential. For STO this is wide and shallow, whereas for ETO it is deep and narrow.[6,7] This implies that dynamically ETO can be classified as "hard" whereas STO is "soft".[7] Furthermore, for ETO the coupling between the transverse optic and the transverse acoustic modes is strongest at the zone boundary opposite to STO where it is strongest near the zone center.[8] The cubic to tetragonal structural phase transition close to room temperature is meanwhile well documented for ETO by various experimental studies.[1,9-14] However, variations of sample quality, local structural disorder,[10] presence of Eu$^{+2}$/Eu$^{+3}$[15], or oxygen vacancies [14] are known to slightly modify the structural phase transition temperature (T$_S$).

In the above mentioned theoretical analysis an additional Heisenberg term has been added to the Hamiltonian to account for the 4f$^7$ related spins, which are coupled to the lattice through a biquadratic coupling term.[16] This term has significant influence on the spin and lattice dynamics since T$_S$ is strongly enhanced by it as compared to STO, and the spins remain active up to temperatures far above T$_N$. The latter fact has the consequence that "hidden" magnetism is present at high temperatures as demonstrated by various experiments[16-18] with the most prominent observation that T$_S$ could be shifted to higher temperature with a magnetic field.[19] Another important result was achieved using the muon spin rotation (μSR) technique, which is a unique local probe to detect any kind of magnetism in a material. Temperature dependent μSR experiments detected that different dynamic stages of magnetism are realized in ETO with coexisting regimes of magnetic and non-magnetic domains.[17] The corresponding onset temperatures are at

$T_S$, $T^* \approx 200K$, and $T' \approx 100K$ including slight precursor effects above $T_S$.[17] The exact characteristics of this hidden magnetism at elevated temperatures are yet unclear, but have to be associated with the coupling between the zone center optic/zone boundary acoustic phonons and a lattice modified para-magnon branch. The interplay between all of these results in the formation of local dynamically fluctuating magnetic nano-regions. These follow the temperature dependence of the soft acoustic mode but remain on the average paramagnetic until $T_N$ is reached and long range antiferromagnetic order sets in.[19]

Here, we present high quality Synchrotron Powder Diffraction (SXPD) data of ETO with in-situ application of small magnetic fields (up to 480mT) in the temperature range 100-300 K. Our study aims to identify magnetic field induced structural changes to clarify the kind of magnetism of the fluctuating domains. Previous micro-Raman experiments performed on the same samples with in situ application of a similar magnetic field revealed the appearance of new modes in the Raman spectrum below ~ 240 K.[20]

High quality 2D synchrotron powder diffraction patterns of a polycrystalline ETO sample have been collected ($\lambda=0.6001\text{Å}$) using the 2M PILATUS detector in the temperature range 100-300K with in-situ application of a static magnetic field (up to 480 mT, Figure S1) at the XRD1 beamline of Elettra Syncrotron Trieste. Details about the sample preparation and characterization can be found in Ref.[21]. The capillary containing the sample was placed above one of the edges of a large Nd magnet where the magnetic field was found to be maximum (left inset in Fig.S1). By varying the distance between the capillary and the magnet the strength of the magnetic field could be tuned. The calibration of the magnetic field strength with respect to the distance from the magnet (Fig. S1, right inset) has been obtained by a magnetometer with exactly the same set up. The 2D diffraction images were converted to $I(2\theta)$ patterns after correcting for distortions and refining the detector distance and orientation using a $LaB_6$ as a standard with the program FIT2d.[22,23] Structural parameters have been obtained by Rietveld refinement using the Fullprof suite.[24] The peak shapes were modeled using the Thompson-Cox-Hastings pseudo-Voigt convoluted with an axial divergence asymmetry function, and the background was estimated by interpolating between selected points, successfully refined. In the last refinement cycles, scale factor, cell parameters, positional coordinates, background points and line profile parameters were allowed to vary.

The diffraction patterns for all applied magnetic fields could be very well fitted by assuming cubic (C) symmetry at 300K (RT) (space group Pm-3m) and tetragonal (T, space group I4/mcm) at temperatures T<290K (Fig.1 and Fig.S2).[10] With decreasing temperature two characteristic features of the pattern indicating the lowering of symmetry become evident for all applied fields, i.e. the splitting or broadening of the $(004)_C$ reflection (Fig.1 and Fig.S2 right insets) and the appearance of the $(211)_T$ superstructure peak (left insets in Fig.1 and Fig.S2). The use of the 2M PILATUS detector enabled us to detect the weak $(211)_T$ superstructure peak even at the strongest applied field (left inset in Fig.S2). The $(400)_C \rightarrow (008)/(440)_T$ peak splitting, characteristic of the Pm-3m$\rightarrow$ I4/mcm symmetry lowering, as well as the superstructure $(211)_T$ reflection are well described by the Rietveld refinement (Fig.1 and Fig.S2). Inspection of the diffraction patterns at 100K shows that the application of a 480mT magnetic field clearly reduces the splitting of the $(004)_C$ reflection peak (right inset Fig.1 and Fig.S2).

At 100K, where the low temperature tetragonal phase is well established diffraction patterns as a function of the applied field have been collected to characterize the evolution of the pseudo-cubic (pc) lattice parameters $a_{pc}=a/\sqrt{2}$ and $c_{pc}=c/2$ (Fig.2) as well as the relative intensity of the weak superstructure $(211)_T$ reflection (associated with the displacements of the O2 atoms) with respect to the unperturbed $(111)_C$. The lattice parameters show a tendency to merge and the relative intensity of the $(211)_T$ reflection to vanish around ~550-600mT (Fig.2). Since we have no XRD data at higher magnetic fields, we cannot decide whether the tetragonal distortion will vanish around this field or a minimum will be achieved to increase again, as other measurements at higher fields (>0.5T) have shown.[17] It is interesting to note that similar threshold field values (0.6-0.7T) have also been reported before.[25-26]

The abrupt increase of $a_{pc}$ as a function of the magnetic field at 100K correlates with the decrease of the relative intensity of the superstructure peak (associated to the oxygen octahedral rotations) (Fig.2), signaling a non-linear dependence of the structural parameters on the magnetic field. The presence of hysteresis was tested by removing the magnetic field at 100 K and taking repeatedly XRD patterns. By monitoring the splitting of the $(400)_C$ reflection an immediate relaxation of the effect with no hysteresis of the lattice response is observed (Fig.S3).

The temperature evolution of structural parameters has been studied for magnetic fields close to and above the threshold value, i.e. 340mT, 439mT, and 480mT, shown in Figure 3 in comparison to those at zero field. Without the magnetic field $c_{pc}$ remains almost independent of temperature within the range 100-300K studied while the $a_{pc}$ lattice parameter almost linearly decreases with decreasing temperature as previously reported,[10,14] though for the highest field applied there are changes in slope around 200K and 250K (Fig.3). The data clearly evidence that the magnetic field strongly affects the lattice of ETO, with this effect being most pronounced in the $a_{pc}$ lattice parameter while $c_{pc}$ is slightly reduced with the application of a magnetic field (Fig.3). In the 100 to 300K temperature range the overall change of the $a_{pc}$ lattice parameter ($\Delta a_{pc}=a_{pc,300K}-a_{pc,100K}$) decreases with increasing field intensity with $\Delta a_{pc}$=0.01075(4)Å, 0.0095(4)Å, 0.00807(4)Å and 0.00584(4)Å for B=0mT, 340mT, 439mT, and 480mT, respectively.

In spite of the reduction of the $c_{pc}$-$a_{pc}$ splitting with the applied field, it is obvious from Fig.3 that for all magnetic fields the temperature where the $c_{pc}$ and $a_{cp}$ start to diverge and the cubic to tetragonal phase transition occurs is close to RT in accordance with the temperature dependence of the $a_{pc}/c_{pc}=\sqrt{2}a/c$ ratio (Fig.S4). Similar results are obtained from the temperature dependence of the relative intensity of the superstructure peak shown in Fig.4 in agreement with previous heat capacity measurements, where $T_S$ increases slightly with the field (approximately 5K with 9T, Ref.[17,19]). For all magnetic fields $a_{pc}$ follows the same trend in the temperature range ~250-300K and only below ~250K deviations set in where the effect for the highest magnetic field is very pronounced (Fig.3). At this field strength and around 200K a further change in slope for $a_{pc}$ takes place (Fig.3).

The tetragonal I4/mcm structure is the consequence of a pairwise out of phase rotation of the oxygen octahedral cage (inset in Fig.S5), with the rotation angle φ (shown in Fig. S5) being the order parameter of the phase transition and calculated using the lattice constants of perfect octahedra according to φ= arccos

(√2a/c). Similar results have been obtained from the equation tan$\varphi$=1-4$x$[O2] where $x$[O2] is the fractional coordinate $x$ of the oxygen O2 atom.[10] The overall value of the order parameter and the $a_{pc}/c_{pc}$ ratio are apparently reduced by the magnetic field. Based on the above observations one may argue that the effect of the magnetic field is to antagonize the tetragonal phase and restore the cubic phase. But for all magnetic fields it is obvious from Fig.4, Figs.S4, S5 that $T_S$ is close to RT and barely affected by the field in agreement with other results using higher magnetic fields.[19]

In the tetragonal phase of ETO there are three distinct Eu-O distances, corresponding to the apical oxygen (Eu-O1) and the planar oxygens O2 split into two (short Eu-O2 and long Eu-O2') caused by the antiferrodistortive rotations of the oxygen octahedra. At 100K the splitting of the planar Eu-O2 bonds has been found to decrease ubruptly at a threshold field of ~400mT (Fig.S6). The temperature dependence of the Eu-O2 distances for different magnetic fields is presented in Fig.S7. The solid lines are polynomial fits confirming that the cubic to tetragonal phase transition is close to RT for all fields, with the progressive lowering of the difference between the Eu-O2' and Eu-O2 being consistent with the decreasing rotation angle of the oxygen octahedra (Fig.S5).

The tetragonal distortion in perovskites can be quantified by using the symmetry-adapted spontaneous strains, $e_a$=(2$e_1$+$e_3$) (volume strain) and $e_t$= 2($e_3$−$e_2$) (tetragonal strain), where $e_1$ and $e_2$ are the strain components calculated using the pseudo-cubic room temperature lattice constants $a_{pc}$ and $c_{pc}$: $e_1 = e_2 = (a_{pc} − a_0)/a_0$ and $e_3 = (c_{pc} − a_0)/a_0$. The temperature dependence of the spontaneous strains provides information of any possible symmetry changes.[14] Fig.5a shows the temperature dependence of the tetragonal spontaneous strain in the presence of the applied magnetic fields. For all applied fields the tetragonal strain approaches zero close to RT as expected and in agreement with previous heat capacity measurements.[19] Our results shows that $e_t$ at specific temperature decreases with increasing external magnetic field. The temperature dependence of the volume strain (Fig.5b) reflects the changes of the lattice constants. It shows that anomalies in it appear with the application of the magnetic field and pronouncedly set in around 250K. A clear change in slope is also observed around 200K at the maximum field (480mT). Anomalies induced by the magnetic field at about the same characteristic temperatures (210K and ~255K) have been reported before and attributed to the appearance of weak ferromagnetism.[17] From our data we conclude that $T_S$ is not affected by the magnetic field and only below ~250K strong changes in the lattice constants with temperature occur.

The lattice responses detected at T<$T_S$ most likely stem from a further reduction in symmetry as concluded from birefringence data on epitaxial thin films of ETO on STO. However, a clear conclusion cannot be drawn from our powder diffraction data since new lines in the diffraction patterns collected in the presence of a magnetic field at 100K could not be detected. On the other hand, the birefringence data revealed an inequivalence of the [100] and [010] directions, which is not compatible with tetragonal symmetry.[27]

Based on the absence of additional diffraction lines at temperatures lower than $T_S$, one may assume that the observed anomalous lattice response to the magnetic field is related to an isostructural lattice transformation. However, in view of the small distortion occurring at the cubic-tetragonal transition and the results from birefringence on thin film[27] more experiments are needed to substantiate this conclusion.

Theoretically, the second nearest neighbor superexchange interaction between Eu 4f spins via the bridging oxygen ions is affected by the spin–lattice coupling in a subtle way.[16] In addition, the direct nearest neighbor antiferromagnetic exchange between Eu-Eu ions in the ab plane and along the c-axis is also sensitive to lattice effects.[16] Both nearest and second nearest Eu-Eu distances (inset in Fig.S5) follow the trends of the respective lattice constants (Fig.3). All structural elements react on the field below $T_s$ at ~250K and around 200K an additional change of slope of the volume strain (Fig.5b) is evident best seen for the highest applied field (480mT), indicative for a further phase transition. By comparing the present data with low temperature magnetostriction results direct conclusions can be drawn, namely an increase in the lattice volume supports ferromagnetic order, whereas a volume contraction takes place for antiferromagnetic order. Figure 6 shows the magnetic field dependence of the relative change of the pseudo cubic unit cell volume ($V_{pc}=V/4$) with the field at various temperatures, $[V_{pc}(T,B)-V_{pc}(T,B=0)]/V_{pc}(T,B=0)$. Obviously, the effect of the field becomes pronounced at 350mT and even more rapidly above ~400mT. It is striking that the magnetostriction changes sign around 200K, namely from expanding to shrinking, signaling a boundary where the magnetic properties change. By comparing this behavior with the one below $T_N$ a tentative conclusion is that intrinsic ferromagnetism changes to intrinsic antiferromagnetism.[28] Apparently weakly antiferromagnetically ordered clusters appear for T>200K whereas below this characteristic temperature ferromagnetism is supported by the field.[29] These conclusions are supported by µSR measurements and results from birefringence where 200K signals a change in the magnetic character together with a structural transition.

In conclusion, we have shown that a small (<0.5T) external magnetic field does not change the cubic to tetragonal structural phase transition temperature $T_S$. On the other hand, the magnetic field has a significant influence on the spontaneous tetragonal strain, the order parameter φ, the unit cell volume, and the magnetostriction. At a threshold field of ~400mT the lattice changes are most pronounced. Two characteristic temperatures (~200K and ~250K), which agree with previous experiments,[17-19] have been identified where the temperature dependence of the lattice parameters are substantially altered. Furthermore, at ~200K the magnetostriction changes sign, thereby in a pronounced manner signaling the modification of the magnetic response, apparently from a crossover between FM and AFM local order.

**Acknowledgements**

We thank J. Köhler of Max Planck Institute for providing the polycrystalline samples and Elettra Sincrotrone in Trieste for beam time allocation.

**Figure Captions**

Figure 1 (color online). Experimental (red crosses), calculated (black solid line) diffraction intensities and their difference (bottom blue solid line) at 100K and no magnetic field. Bars indicate Bragg positions of the I4/mcm space group ($R_{wp}$=9.6%). Inset show the region near the $(211)_T$ superstructure peak (left) and the $(008)_T$ $(440)_T$ peaks at 0mT (red crosses) and 480mT (magenta crosses) respectively (right).

Figure 2 (color online). The dependence on the applied magnetic field at 100 K of the pseudo-cubic lattice parameters ($c_{pc}$, blue triangles, $a_{pc}$ black spots) and the relative intensity of the superstructure $(211)_T$ to the $(111)_C$ peak (red squares). Continuous lines are guides to the eye.

Figure 3 (color online). Evolution of the pseudocubic lattice parameters upon temperature for different applied magnetic fields, B=0 (black circles), 340 (blue squares), 439 (green stars), and 480mT (red trianges). Open symbols correspond to $a_{pc}$ and closed symbols to $c_{pc}$ respectively.

Figure 4 (color online). Temperature dependence of the relative intensity of the $(211)_T$ superstructure peak to the intensity of the $(111)_C$ peak for B=0 (black circles), 439 (green stars), and 480mT (red triangles). The continuous lines are polynomial fits to data.

Figure 5 (color online). Temperature dependence of the spontaneous tetragonal strain $e_t$ (a) and the spontaneous volume strain $e_V$ (b) for different magnetic fields.

Figure 6 (color online). Magnetic field dependence of the relative unit cell volume (compared to its value at the same temperature without a field) for various temperatures. Lines are guide to the eye.

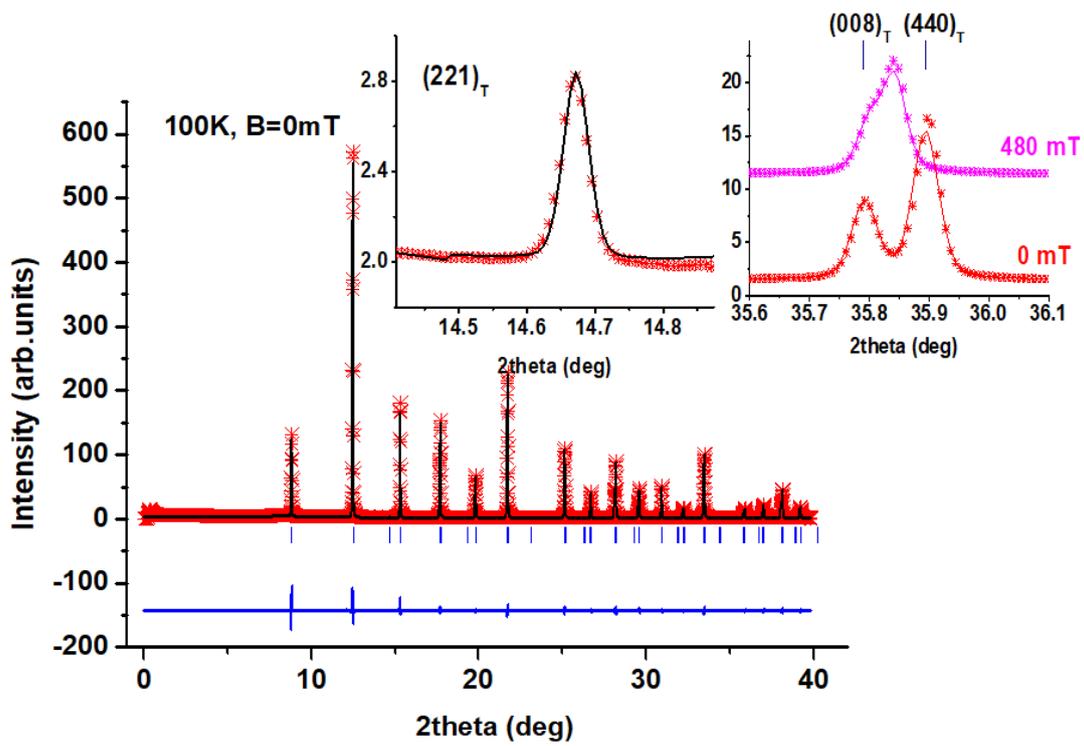

Figure 1.

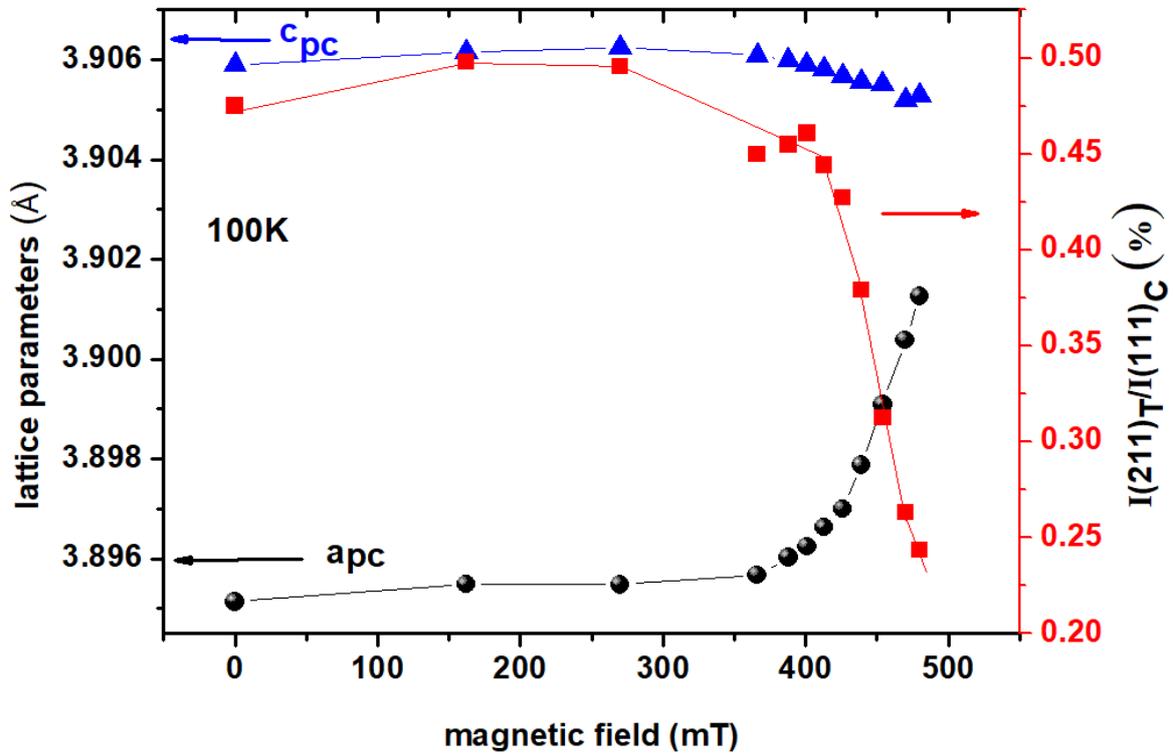

Figure 2.

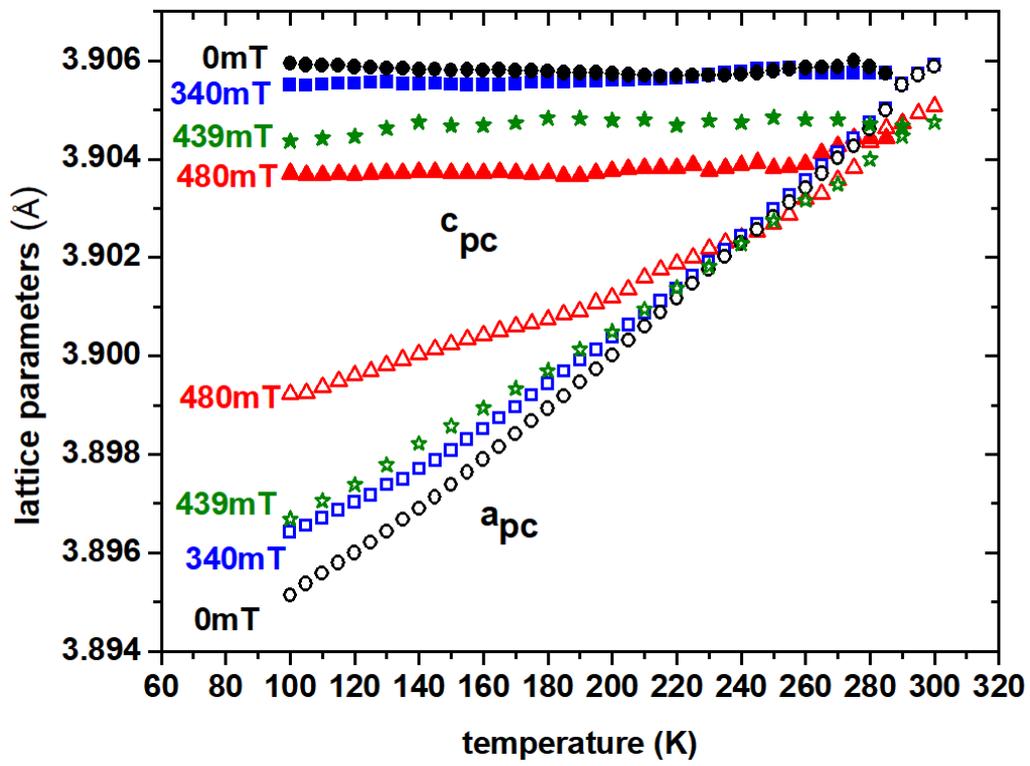

Figure 3.

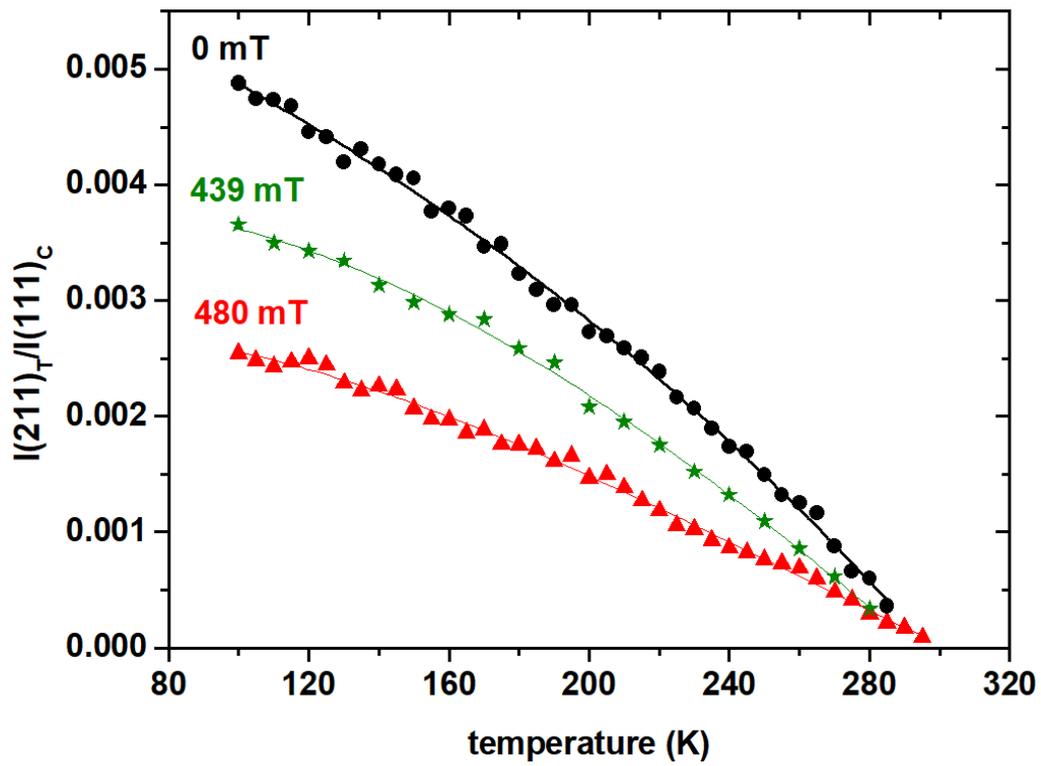

Figure 4.

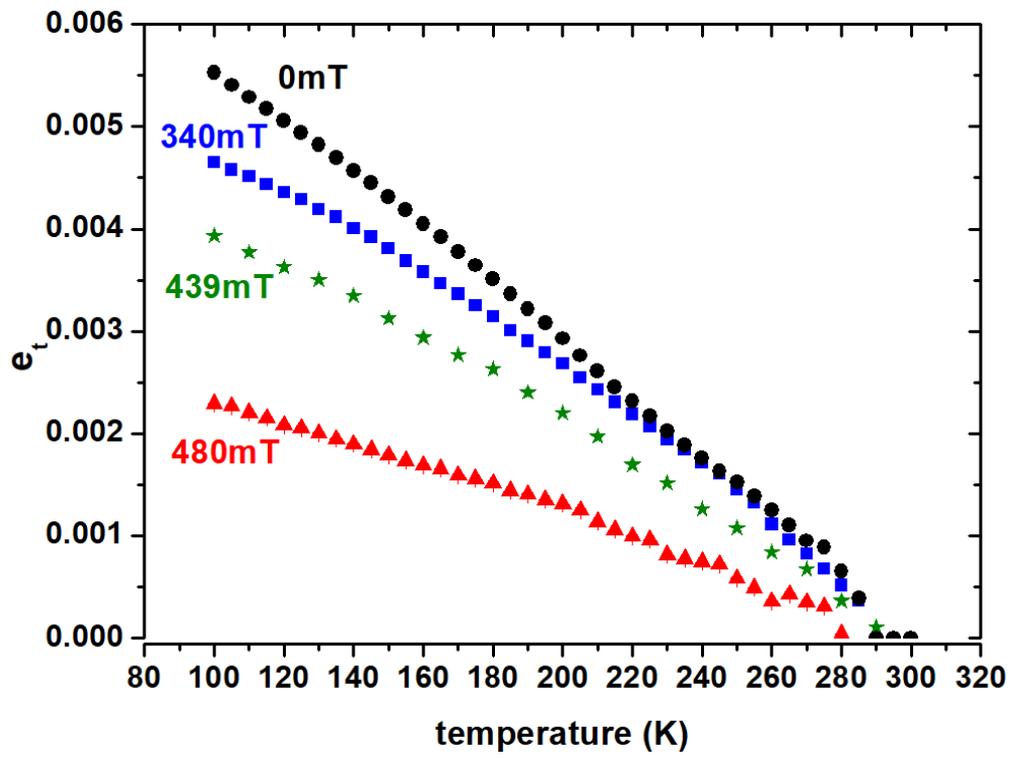

Figure 5a.

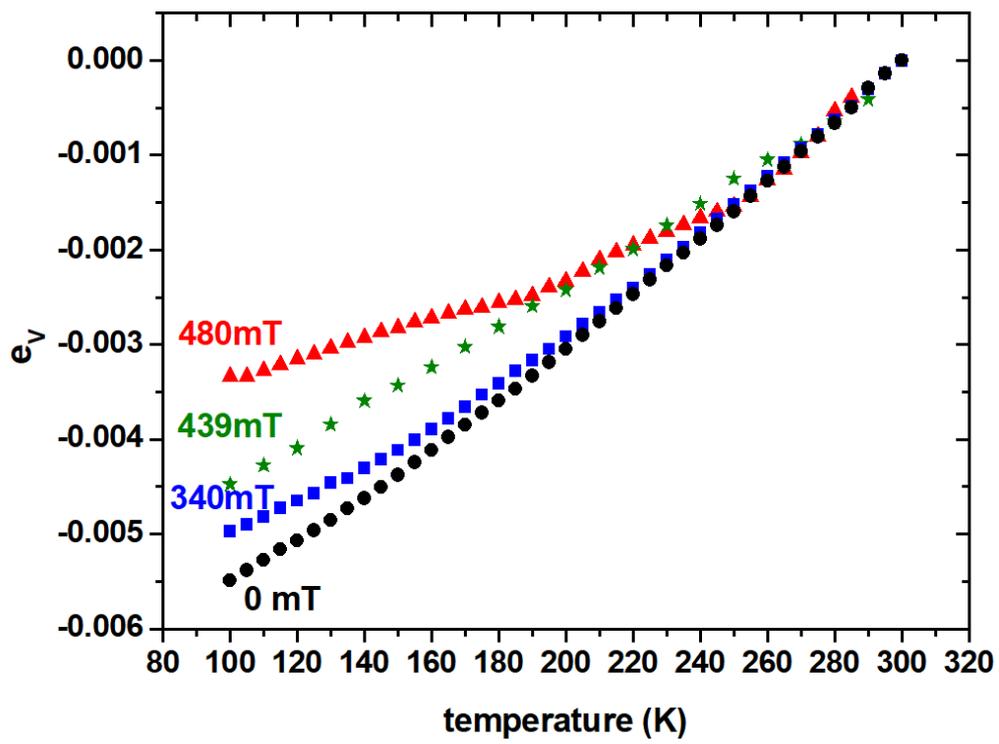

Figure 5b.

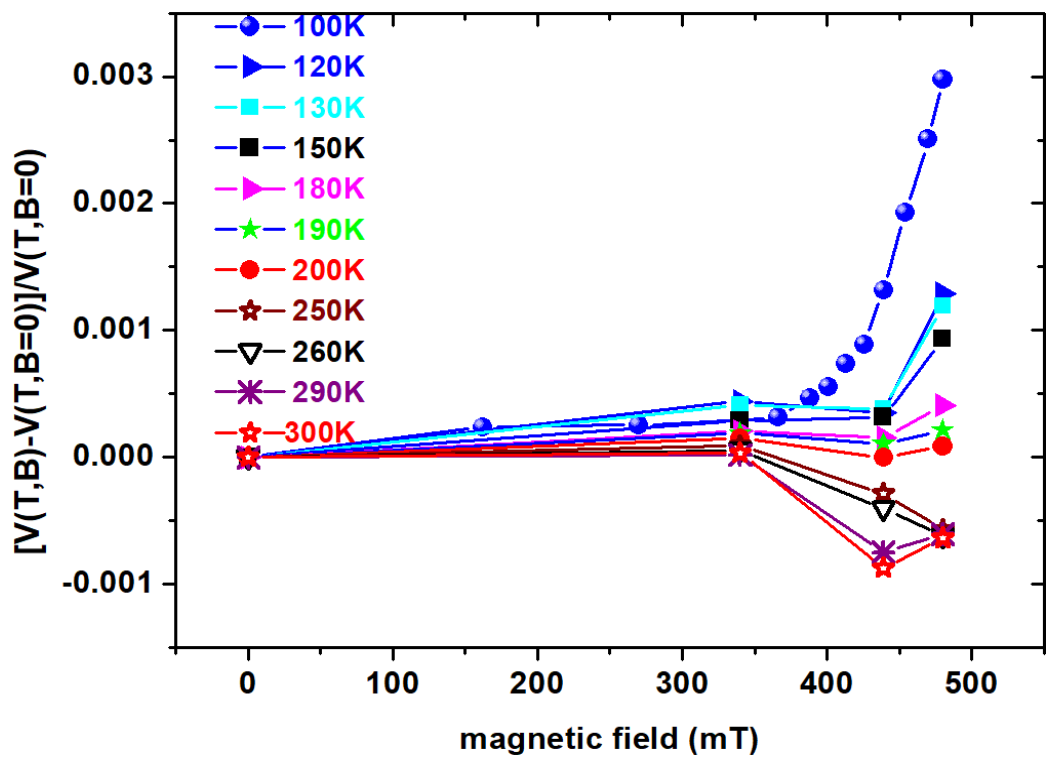

Figure 6.

## Supplementary Material

The capillary (0.2mm diameter) was placed above and parallel to one of the edges of the Nd magnet where the magnetic field was stronger (Figure S1). By varying the distance between the capillary and the magnet the strength of the magnetic field could be tuned. At a constant temperature 100K diffraction patterns have been collected as a function of the applied field (0-480mT). The calibration of the magnetic field strength with respect to distance from the magnet (Figure S1 inset) has been obtained by a magnetometer (HEME fluxmeter Type T2B) with exactly the same set up, substituting the capillary with the instrument probe. The Oxford CryoSystems cryo-stream cooler has been used for the low temperature measurements by warming from 100K to 300K with 10K or 5K steps. To avoid averaging out any possible effect of the magnetic field on the lattice we chose to keep the sample still during data acquisition. In order to increase the statistics of our data and exclude any effect from homogeneity, we collected at each temperature or field three diffraction patterns from three different sample point (200 micron apart) on the capillary, which were then averaged.

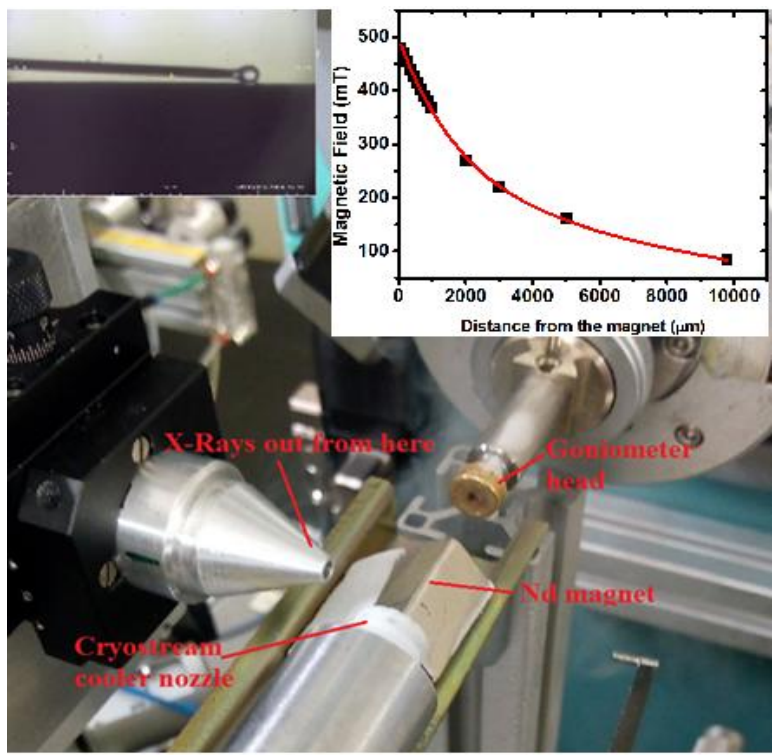

Figure S1 (color online). Experimental setup for the in-situ application of a magnetic field. Left inset shows the capillary located above the magnet as seen from the on-axis CCD camera (view of the incoming X-rays). Right inset presents the calibration curve of the magnetic field strength as a function of the distance from the magnet.

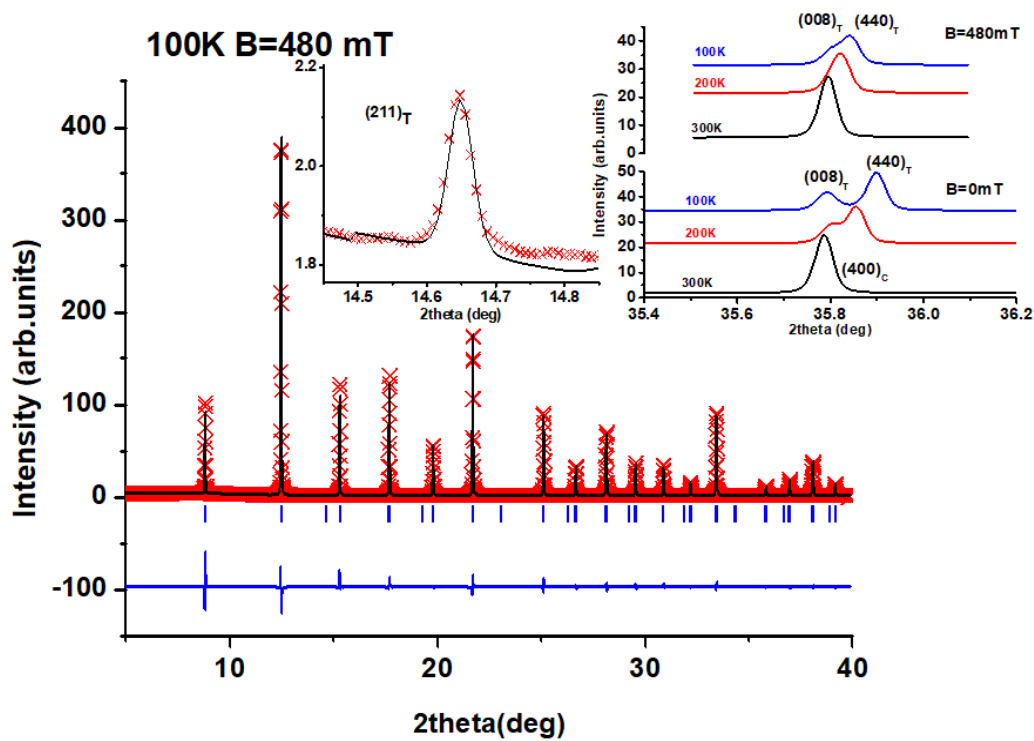

Figure S2 (color online). Experimental (red crosses), calculated (black solid line) diffraction intensities and their difference (bottom blue solid line) at 100K and 480mT. Bars indicate Bragg positions of the I4/mcm space group ($R_{wp}$=10.6%). Insets show the region near the $(211)_T$ superstructure peak at $2\theta=14.46°$ (up left) and the $(008)_T$ and $(440)_T$ reflections at various temperatures for B=0mT and B=480mT (up right).

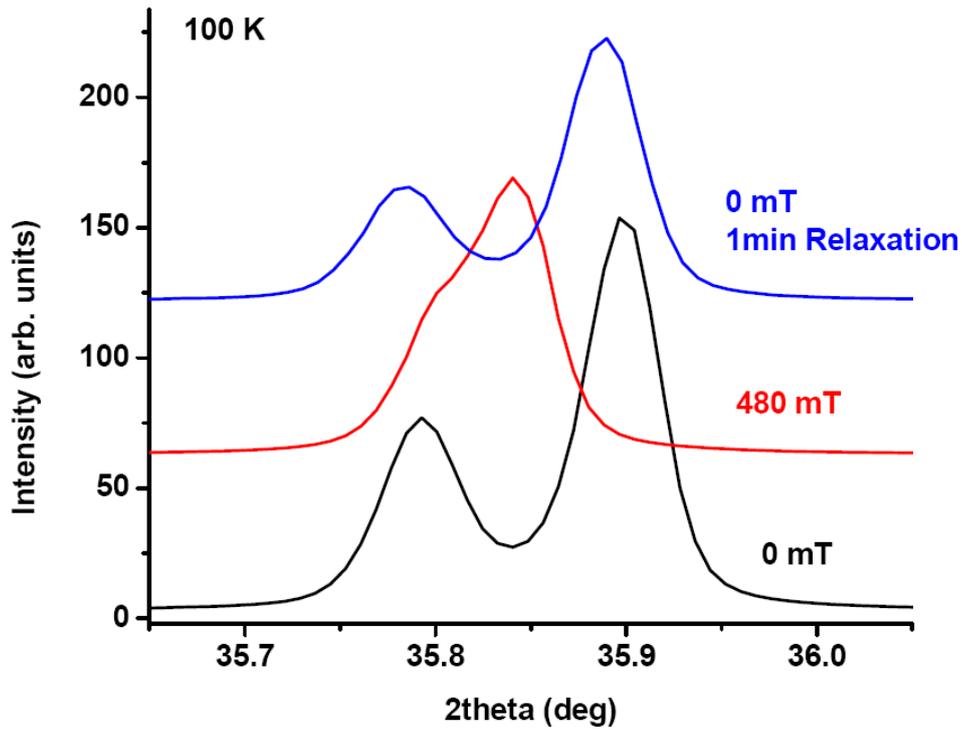

Figure S3 (color online). Splitting of the $(400)_C$ peak measured at 100K in-situ with zero magnetic field (bottom, black), applying a field of 480mT (middle, red) and 1min after the removal of the magnetic field (top, blue).

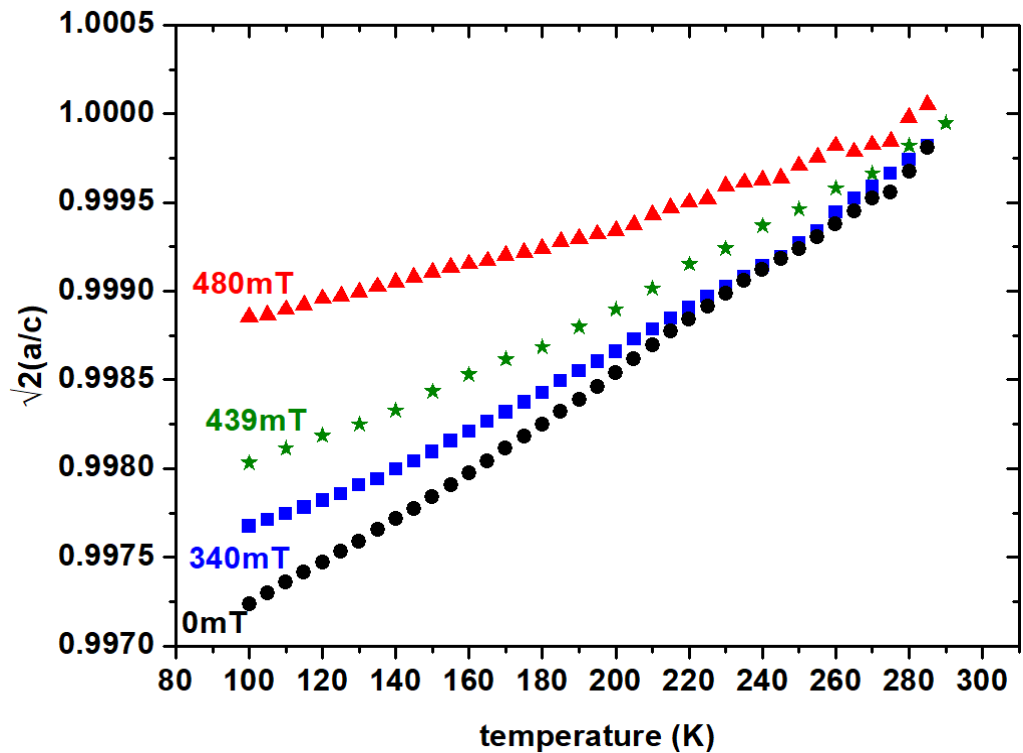

Figure S4 (color online). The thermal dependence of the tetragonal $a_{pc}/c_{pc}$ ratio for different applied fields 480mT (red triangles), 439mT (green stars), 340mT (blue squares), and 0mT (black spots).

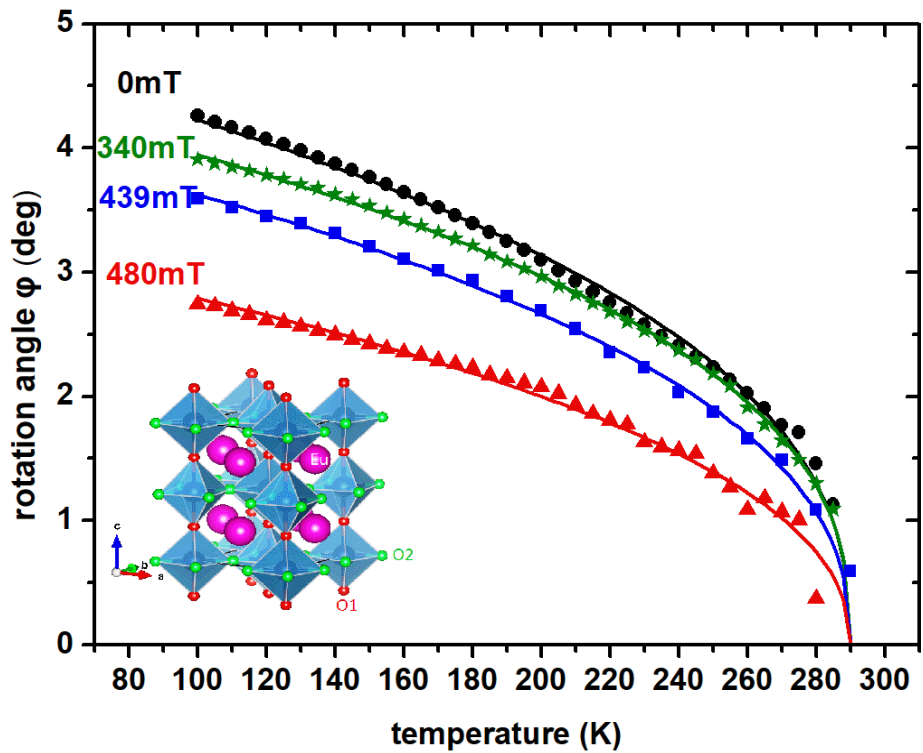

Figure S5 (color online). Temperature dependence of the oxygen octahedral rotation $\varphi$. Solid lines are fits to the equation $\varphi=A(T-T_s)^\beta$ with $T_s=290$K fixed and $\beta=0.400\pm0.006$, $A=0.52\pm0.01$ (0mT, black spots), $\beta=0.381\pm0.002$, $A=0.53\pm0.008$ (340mT, green stars), $\beta=0.41\pm0.02$, $A=0.41\pm0.4$ (439mT, blue squares), and $\beta=0.44\pm0.01$, $A=0.27\pm0.01$ (480mT, red triangles). Inset shows the structure with the rotation angle of the $TiO_6$ octahedra exaggerated.

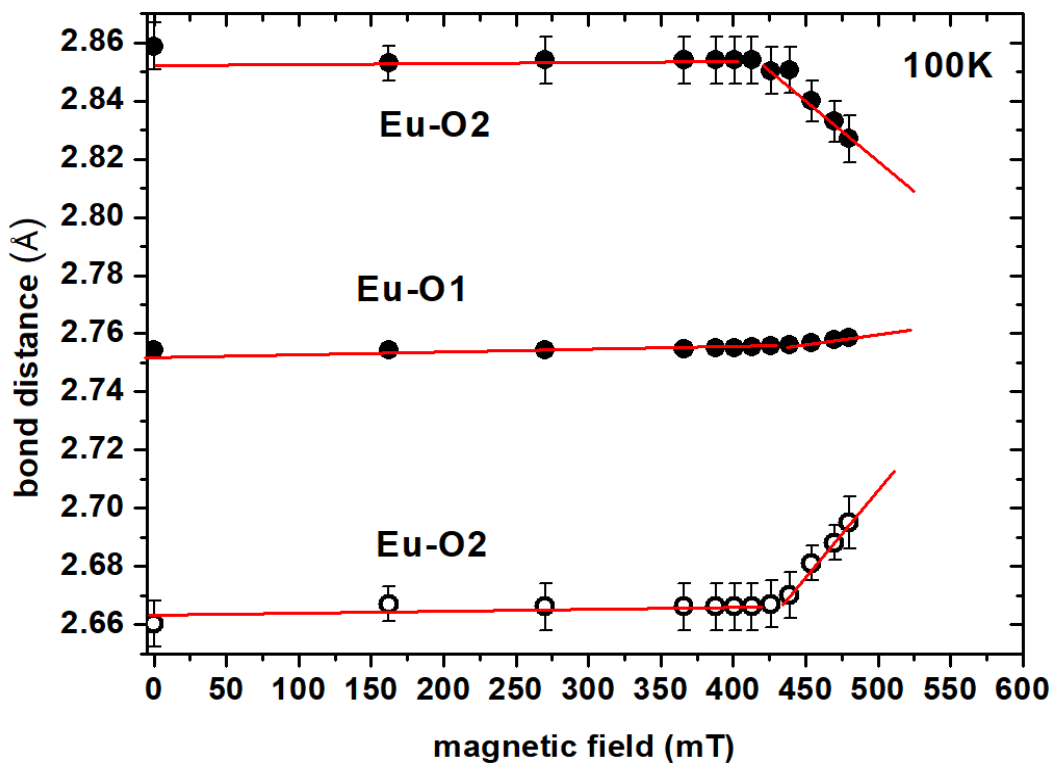

Figure S6 (color online). Eu-O distances as a function of the applied magnetic field at 100 K. Continuous lines are guides to the eye. Error bars for Eu-O1 are smaller than the spots size.

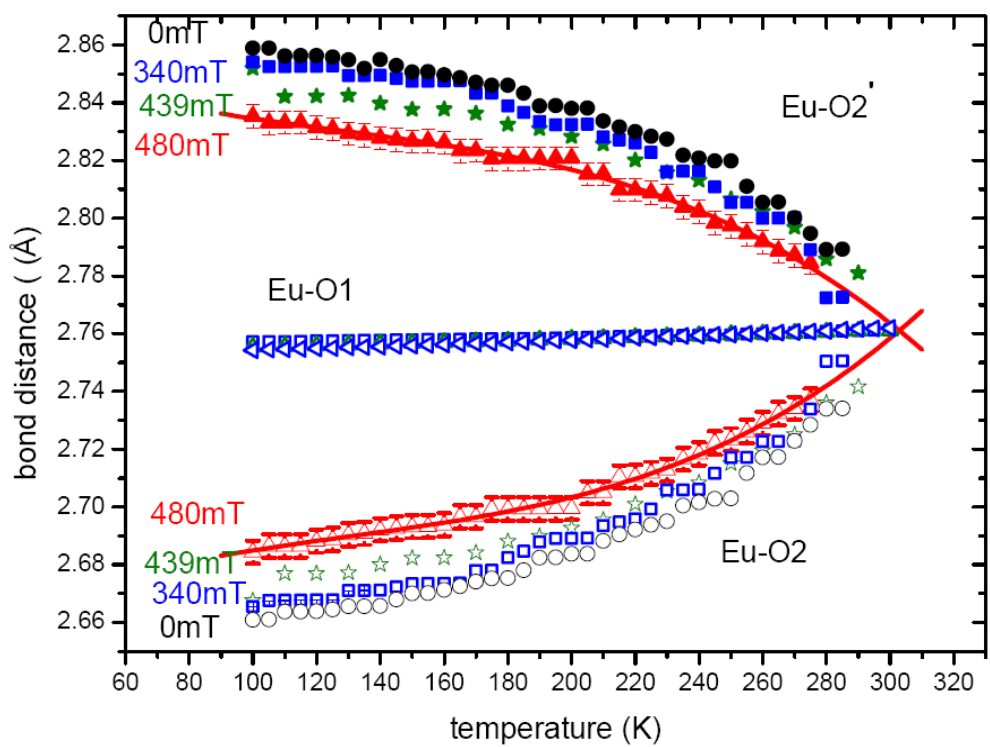

Figure S7 (color online). Temperature dependence of the Eu-O bonds for various magnetic fields, 480mT (black spots), 439mT (green stars), 340mT (blue squares), and 0mT (black spots). Solid lines are polynomial fits. Error bars of the Eu-O1 bonds are smaller than the symbols.